\documentclass[aps,preprint]{revtex4}
\usepackage{graphicx,color}
\usepackage{psfrag}
\usepackage{amssymb}
\usepackage{eso-pic}
\usepackage[dvips,letterpaper]{geometry} 
\usepackage{amsmath}

\newcommand\BackgroundPicture{%
   \put(0,0){%
     \parbox[b][\paperheight]{\paperwidth}{%
       \vfill
       \centering
       \includegraphics[width=\paperwidth,height=\paperheight,%
                        keepaspectratio]{RealLobster.eps}%
       \vfill
     }}}  

\AddToShipoutPicture{\BackgroundPicture}    

\ClearShipoutPicture                    

\begin{document}

%
%
%
%
\def\oti{{\otimes}}
\def\lb{ \left[ }
\def\rb{ \right]  }
\def\tilde{\widetilde}
\def\bar{\overline}
\def\hat{\widehat}
\def\*{\star}
\def\[{\left[}
\def\]{\right]}
\def\({\left(}		\def\BL{\Bigr(}
\def\){\right)}		\def\BR{\Bigr)}
	\def\BBL{\lb}
	\def\BBR{\rb}
%
%
\def\zb{{\bar{z} }}
\def\zbar{{\bar{z} }}
\def\frac#1#2{{#1 \over #2}}
\def\inv#1{{1 \over #1}}
\def\half{{1 \over 2}}
\def\d{\partial}
\def\der#1{{\partial \over \partial #1}}
\def\dd#1#2{{\partial #1 \over \partial #2}}
\def\vev#1{\langle #1 \rangle}
\def\ket#1{ | #1 \rangle}
\def\rvac{\hbox{$\vert 0\rangle$}}
\def\lvac{\hbox{$\langle 0 \vert $}}
\def\2pi{\hbox{$2\pi i$}}
\def\e#1{{\rm e}^{^{\textstyle #1}}}
\def\grad#1{\,\nabla\!_{{#1}}\,}
\def\dsl{\raise.15ex\hbox{/}\kern-.57em\partial}
\def\Dsl{\,\raise.15ex\hbox{/}\mkern-.13.5mu D}
%
%
\def\ga{\gamma}		\def\Ga{\Gamma}
\def\be{\beta}
\def\al{\alpha}
\def\ep{\epsilon}
\def\vep{\varepsilon}
\def\la{g}	\def\La{\Lambda}
\def\de{\delta}		\def\De{\Delta}
\def\om{\omega}		\def\Om{\Omega}
\def\sig{\sigma}	\def\Sig{\Sigma}
\def\vphi{\varphi}
%
%
\def\CA{{\cal A}}	\def\CB{{\cal B}}	\def\CC{{\cal C}}
\def\CD{{\cal D}}	\def\CE{{\cal E}}	\def\CF{{\cal F}}
\def\CG{{\cal G}}	\def\CH{{\cal H}}	\def\CI{{\cal J}}
\def\CJ{{\cal J}}	\def\CK{{\cal K}}	\def\CL{{\cal L}}
\def\CM{{\cal M}}	\def\CN{{\cal N}}	\def\CO{{\cal O}}
\def\CP{{\cal P}}	\def\CQ{{\cal Q}}	\def\CR{{\cal R}}
\def\CS{{\cal S}}	\def\CT{{\cal T}}	\def\CU{{\cal U}}
\def\CV{{\cal V}}	\def\CW{{\cal W}}	\def\CX{{\cal X}}
\def\CY{{\cal Y}}	\def\CZ{{\cal Z}}

\def\rvac{\hbox{$\vert 0\rangle$}}
\def\lvac{\hbox{$\langle 0 \vert $}}
\def\comm#1#2{ \BBL\ #1\ ,\ #2 \BBR }
\def\2pi{\hbox{$2\pi i$}}
\def\e#1{{\rm e}^{^{\textstyle #1}}}
\def\grad#1{\,\nabla\!_{{#1}}\,}
\def\dsl{\raise.15ex\hbox{/}\kern-.57em\partial}
\def\Dsl{\,\raise.15ex\hbox{/}\mkern-.13.5mu D}
%
%
%
\font\numbers=cmss12
\font\upright=cmu10 scaled\magstep1
\def\stroke{\vrule height8pt width0.4pt depth-0.1pt}
\def\topfleck{\vrule height8pt width0.5pt depth-5.9pt}
\def\botfleck{\vrule height2pt width0.5pt depth0.1pt}
\def\Zmath{\vcenter{\hbox{\numbers\rlap{\rlap{Z}\kern
0.8pt\topfleck}\kern 2.2pt
                   \rlap Z\kern 6pt\botfleck\kern 1pt}}}
\def\Qmath{\vcenter{\hbox{\upright\rlap{\rlap{Q}\kern
                   3.8pt\stroke}\phantom{Q}}}}
\def\Nmath{\vcenter{\hbox{\upright\rlap{I}\kern 1.7pt N}}}
\def\Cmath{\vcenter{\hbox{\upright\rlap{\rlap{C}\kern
                   3.8pt\stroke}\phantom{C}}}}
\def\Rmath{\vcenter{\hbox{\upright\rlap{I}\kern 1.7pt R}}}
\def\Z{\ifmmode\Zmath\else$\Zmath$\fi}
\def\Q{\ifmmode\Qmath\else$\Qmath$\fi}
\def\N{\ifmmode\Nmath\else$\Nmath$\fi}
\def\C{\ifmmode\Cmath\else$\Cmath$\fi}
\def\R{\ifmmode\Rmath\else$\Rmath$\fi}

\def\barray{\begin{eqnarray}}
\def\earray{\end{eqnarray}}
\def\beq{\begin{equation}}
\def\eeq{\end{equation}}

\def\no{\noindent}

\title{Critical point of the two-dimensional Bose gas: an
S-matrix approach}
\author{Pye-Ton How and Andr\'e  LeClair}
\affiliation{Newman Laboratory, Cornell University, Ithaca, NY} 
\date{May  2009}

\bigskip\bigskip\bigskip\bigskip

\begin{abstract}

A new treatment of the critical point of the two-dimensional 
interacting Bose gas is presented.  In the lowest order
approximation we obtain the critical temperature 
$T_c \approx 2 \pi n/[ m \log (2\pi/mg)]$,  where $n$ is the density,
$m$ the mass, and $g$ the coupling.  This result is based 
on a new formulation of interacting gases at finite density
and temperature which is reminiscent of the thermodynamic
Bethe ansatz in one dimension.   In this formalism,  the basic
thermodynamic quantities are expressed in terms of a pseudo-energy.
Consistent resummation of 2-body scattering leads to an integral equation
for the pseudo-energy with a kernel based on the logarithm of
the exact 2-body S-matrix.

\end{abstract}


\maketitle

\def\om#1{\omega_{#1}}

\def\Tr{\rm Tr} 
\def\free{\CF} 
\def\xvec{{\bf x}}
\def\kvec{{\bf k}}
\def\kvecp{{\bf k'}}
\def\omk{\om{\kvec}} 
\def\dk#1{(d\kvec_{#1})}
\def\2pid{(2\pi)^d}
\def\fill{{f}}
\def\dkline#1{\underline{\underline{d#1}}}
\def\ket#1{|#1 \rangle}
\def\bra#1{\langle #1 |}
\def\vol{V}
\def\Uhat{\hat{W}}
\def\vhat{\hat{w}} 
\def\bfN#1{{\bf{#1}}}
\def\ketbf#1{|{\bf #1}\rangle}
\def\brabf#1{\langle {\bf #1} | }
\def\adag{a^\dagger}
\def\rme{{\rm e}}
\def\Im{{\rm Im}}
\def\Fhat{\digamma}
\def\dkonly{(d\kvec)} 
\def\U{U}

\section{Introduction}

The properties of interacting Bose gases can be very different
depending on the spatial dimension.    This has become especially
interesting in recent years due to the possibility of experimentally
realizing lower dimensional cold gases  with magnetic and optical traps.   
In 3 dimensions there is a critical point for Bose-Einstein 
condensation (BEC) even in the non-interacting theory, 
with a critical temperature $T_c = 4\pi \( n/\zeta (3/2) \)^{2/3}$
where $n$ is the density and $\zeta$ Riemann's zeta function.  
In two dimensions the same formula becomes 
$T_c = 4\pi n /\zeta (1)$ and since $\zeta (1)$ diverges there is
no critical point at finite temperature for the non-interacting gas.  
It is believed however that the  two-dimensional interacting gas  
has a critical point in the universality class of the Kosterlitz-Thouless
transition  rather than  BEC\cite{Popov,Fisher,Baym,Prokofev}.  
(For a review see \cite{Posaz}.)
This transition has recently been observed in experiments\cite{Kruger}.

In the one-dimensional case the particles effectively behave 
as fermions, the model is integrable, 
 and an exact solution is known based on the thermodynamic 
Bethe ansatz (TBA)\cite{Lieb,YangYang}.
The one-dimensional case illustrates the importance of non-perturbative
methods like the TBA for understanding the physics.  Whereas 
the usual perturbative finite-temperature Feynman diagram method involving
Matsubara frequencies in loops is a standard approach which entangles
zero temperature perturbation theory with quantum statistical sums,   the
TBA represents an entirely different organization of the free energy 
which essentially disentangles the two.  
   More specifically,  the only
property of the model it is based on is the exact  2-body scattering matrix
computed to all orders in perturbation theory at zero temperature.   
In principle a TBA-like organization of the free energy is possible
for non-integrable theories in any dimension based on the formula in
\cite{Ma} which expresses the partition function in terms of the 
exact S-matrix.   The derivation of the TBA as given by Yang and Yang
however was not based  on the result in \cite{Ma} but rather relied 
on the factorizability of the multi-particle S-matrix into 2-body
S-matrices for integrable systems.   For non-integrable systems 
the formalism in \cite{Ma} can be quite complicated since  the free
energy contains N-body terms which do not factorize.   Nevertheless, 
for a dilute gas the consistent resummation of the 2-body scattering
terms can represent a useful approximation that is intrinsically different
from other methods.    

In this work we derive explicit expressions for  the free-energy, 
occupation numbers, etc.,  in this 2-body approximation for non-integrable
models in any spatial dimension.     The result
is TBA-like:   everything is expressed in terms of a pseudo-energy 
that satisfies an integral equation with a momentum-dependent 
kernel which is related to 
a matrix element of the logarithm of the S-matrix.   Our analysis 
builds on the previous work\cite{Leclair},  and the final result 
presented here contains several important technical improvements.

Most of this paper is devoted to developing the formalism in 
generality.  In section III  we describe contributions to the
free energy 
with diagrams, not to be confused with finite temperature Feynman
diagrams,  where the vertices represent $N$-body interactions
for any $N$.   The cluster decomposition of the S-matrix is necessary
to establish the extensivity of the free energy.   In section IV we
derive an integral equation for the occupation numbers (filling
fractions) from a variational principle based on a Legendre transformation
that exchanges the chemical potential with the filling fraction.   
In section V we present the integral equation that consistently resums
the infinite number of 2-body diagrams.   In section VI   the 2-body 
kernel that appears in the integral equation is derived for
interacting Bose gases in 2 and 3 dimensions.    In section VII we compare our approximation 
to the exact TBA for the 1-dimensional case.  

This new formalism is illustrated in the 2-dimensional case.  Since 
the 2-body approximation is reasonably simple in its final form, 
we present a self-contained analysis of the critical point of 
the Bose gas  in the next section, where 
we derive an expression for the coupling constant dependent
critical density.

\section{Critical point of the two-dimensional Bose gas}

In this section we illustrate the main ingredients of our formalism
by applying it to the critical point of the two-dimensional Bose gas. 
Our treatment is significantly different from   previous
 ones\cite{Popov,Fisher,Baym,Prokofev}.    Although we find  results
that are similar to the  known results,  
we find our treatment 
to be  considerably simpler and transparent and doesn't rely
on an effective description of vortices.   
    As will become clear,  the full power of our
formalism is actually unnecessary to obtain these results when the coupling
is very small.  We set $\hbar = k_B = 1$ unless otherwise indicated.

The interacting Bose gas in 2 spatial dimensions is defined by the
following hamiltonian:
\beq
\label{ham.1}
H = \int d^2 \xvec \( \inv{2m} \vert \vec{\nabla} \phi (x) \vert^2 
+ 
\frac{g}{4} \vert \phi(x) \vert^4 \) 
\eeq
where $\phi$ is a complex field and $m$ is the mass of the particles. 
In two dimensions,  the combination $mg/\hbar^2 $ is dimensionless;
in the experiment \cite{Kruger} it is approximately $0.13$.

In the formalism developed in the sequel,  the occupation numbers,
or filling fractions,  are expressed in terms of a pseudo-energy 
$\epsilon (\kvec)$, and the density has the following form:
\beq
\label{crit.1}
n  =  \int \frac{d^2 \kvec}{(2\pi)^2}    \inv{ e^{\beta \epsilon (\kvec)} -1 }
\eeq
where $\beta$ is the inverse temperature $T$.   In an approximation
that consistently resums all the 2-body interactions,  the
pseudo-energy satisfies the integral equation in eq. (\ref{pseudo.energy})
where the kernel $G_2 (\kvec , \kvec' )$ is related to a logarithm 
of the exact 2-body scattering matrix and given explicitly in 
eq. (\ref{kernel.2d}), for convenience reproduced here:
\beq
\label{2dkernel}
G_2 (\kvec_1 , \kvec_2 ) =
- \frac{8}{m} \arctan\(  \frac{mg/8}{1 +  \frac{mg}{8\pi}  \log \( 
\frac{4 \Lambda^2}{|\kvec_1 - \kvec_2 |^2} \)  } \) 
\eeq
where $\Lambda$ is an ultra-violet cut-off.  
    When the coupling $g$ is small,  the kernel is small
and the eq. (\ref{pseudo.energy}) can be approximated by: 
\beq
\label{crit.2}
\epsilon (\kvec ) = \omega_\kvec - \mu  -   \int  \frac{d^2 \kvec'}{(2\pi)^2}  \, 
G_2 (\kvec , \kvec' ) \frac{e^{\beta (\epsilon_{\kvec'} - \omega_{\kvec'} + \mu)}}
{e^{\beta \epsilon_{\kvec'} } -1 }
\eeq
where $\mu$ is the chemical potential and $\omega_\kvec = \kvec^2 /2m $ is the single particle energy. 

Since when $g=0$, $\epsilon (\kvec) =  \omega_\kvec - \mu$,  
an approximate solution to (\ref{crit.2}) is obtained by substituting $\epsilon (\kvec) = \omega_\kvec - \mu$
on the right hand side.   As in Bose-Einstein condensation, 
 the critical point is defined to occur at the value of the
chemical potential $\mu_c$ where the occupation number 
at zero momentum diverges, i.e. $\epsilon (\kvec =0) = 0$. 
This gives the following integral equation for $\mu_c$:
\beq
\label{crit.3}
\mu_c  =  -   \int  \frac{d^2 \kvec}{(2\pi)^2}  \,  G_2 (0, \kvec) \inv {e^{\beta(\omega_\kvec - \mu_c )} -1 }
\eeq

At very weak coupling, to leading order we 
 can neglect the momentum dependence of the kernel and 
simply take $G_2 = -g$.   Consider first the case where the interaction is attractive, 
i.e. $g<0$.  The equation (\ref{crit.3}) becomes
\beq
\label{crit.4}
\beta \mu_c  = - \frac{mg}{2 \pi}  \log \( 1 - e^{\beta \mu_c } \)
\eeq
For negative $g$, the above equation has negative $\mu_c$ solutions, which 
makes physical sense since energy is released when a particle is added.  
Here the attractive interaction makes a critical point possible at finite
temperature and density.  The critical density $n_c \approx \mu_c/g$.   

\begin{figure}[htb] 
\begin{center}
\hspace{-15mm}
\psfrag{Y}{RHS, LHS}
\psfrag{X}{$\beta \mu_c$} 
\psfrag{xc}{$\epsilon_0 = \epsilon_c$}
\psfrag{xo}{$\beta \epsilon_0 = 0.9$}
\includegraphics[width=10cm]{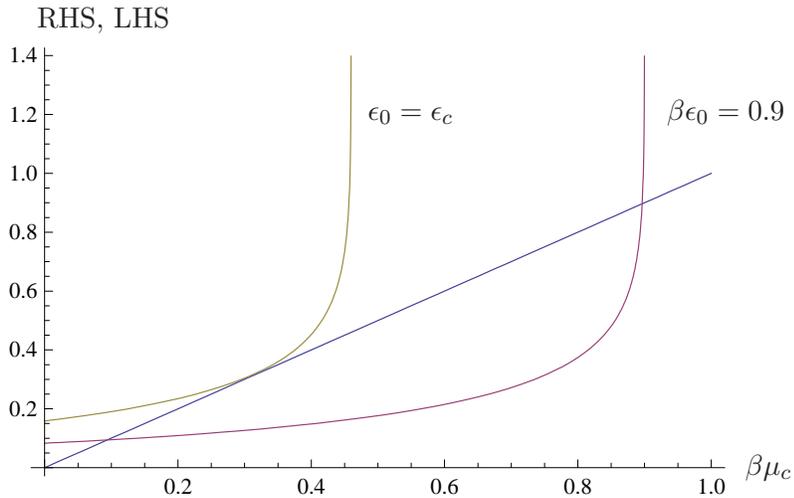} 
\end{center}
\caption{Left and right hand sides of eq. (\ref{crit.5}) as a function of $\beta \mu_c$ 
at two values of the infra-red cut-off $\epsilon_0$, one being the
critical value $\epsilon_c$.}  
\vspace{-2mm}
\label{mucrit} 
\end{figure}

Suppose now the attraction is repulsive,  i.e. $g>0$.   The equation (\ref{crit.4}) 
now has no real solutions for $\mu_c$.  This can be traced to the  infra-red divergence 
of the integral when $\mu_c$ changes sign.    Let us therefore introduce a low-momentum 
cut-off $k_0$, i.e. restrict the integral to $|\kvec| > k_0$.   The equation (\ref{crit.3}) 
now becomes
\beq
\label{crit.5}
\beta \mu_c  =  \frac{mg}{2\pi}  \[  \beta \epsilon_0 -  \beta \mu_c  - 
\log \( e^{\beta (\epsilon_0 - \mu_c )} -1 \) \] 
\eeq
where $\epsilon_0 = k_0^2/2m$.     The above equation now has solutions at positive $\mu_c$,
however for arbitary $k_0$ there are in general two solutions.   If $k_0$ is
chosen appropriately to equal a critical value $k_c$,  then there is a unique solution  $\mu_c$.
This is shown in Figure \ref{mucrit},  where the right and left hand sides of equation (\ref{crit.5}) 
are plotted against $\mu_c$ for two values of $k_0$, one being the
critical value $k_c$.      
Since the slopes of the right and left hand  sides of eq. (\ref{crit.5}) are equal at the critical point,
taking the derivative with respect to $\mu_c$ of both sides gives
the additional equation:
\beq
\label{crit.6}
\beta (\epsilon_c - \mu_c )  =   \log \( 1 +  \frac{mg}{2\pi} \) 
\eeq
where $\epsilon_c = k_c^2 / 2m$.   The two equations (\ref{crit.5}, \ref{crit.6}) determine $\mu_c$
and $\epsilon_c$:  
\barray
\label{crit.7}
\mu_c  &=&  \frac{mgT}{2\pi}  \log \( 1 + \frac{2\pi}{mg} \) 
\\  
\epsilon_c &=& \frac{k_c^2}{2m}  =  T \( 1 + \frac{mg}{2\pi} \) \log \( 1 + \frac{mg}{2\pi} \)  
+ \frac{mgT}{2\pi} \log \(  \frac{2\pi}{mg} \) 
\earray
Note that $\epsilon_c$ goes to zero as $g$ goes to zero.
Finally, making the approximation $\epsilon (\kvec) \approx \omega_\kvec -\mu_c$ in 
eq.  (\ref{crit.1})  and using eq. (\ref{crit.5}) 
one obtains the critical density
$n_c \approx \mu_c /g$:
\beq
\label{crit.8}
n_c \approx \frac{m k_B T}{2\pi \hbar^2 } \log \( 1 + \frac{2\pi}{mg} \) 
\eeq
The above can be expressed as $n_c \lambda^2 = \log ( 1 + 2\pi/mg )$
where $\lambda = \hbar \sqrt{2\pi/mk_BT}$ is the thermal wavelength. 
As expected,  at fixed density,  the critical density 
 goes to infinity as $g$ goes to zero.

In obtaining the above result we have only kept the leading term
in the kernel $G_2$ at small coupling,  which is momentum-independent.  
Systematic corrections to the above results may be obtained by
keeping higher order terms in the kernel.   However since the kernel involves
$\log \kvec^2$,  at small coupling these corrections are quite small,
as we have verified numerically.

In numerical simulations of the critical point,  it was found that 
$\mu_c  \sim  \frac{mgT}{2\pi} \log (2C /mg )$  where $C$ is a constant\cite{Prokofev}.
This form 
agrees with our result (\ref{crit.7}) in the limit of small $g$,  however,  whereas we find
$C = \pi$,  numerically one finds a significantly  higher value $C\approx 13.2$. 
Assuming that $C$ was not overestimated numerically, 
one possible explanation for this discrepancy is based on the renormalization group.  The beta-function for
$g$ can be found by requiring that the kernel $G_2$ in eq. (\ref{kernel.2d}) be independent of
the ultra-violet cut-off $\Lambda$.   To lowest order this gives:
\beq
\label{RG.1}
\Lambda  \frac{dg}{d\Lambda}  =  \frac{m g^2}{4\pi}  
\eeq
which implies the coupling decreases at low energies.  
Integrating this equation leads to the cut-off dependent coupling $g(\Lambda)$:
\beq
\label{RG.2}
\frac{2 \pi}{ m g (\Lambda) }  =  \frac{2\pi}{m g_0} \(  1 +  \frac{mg_0}{4\pi} \log (\Lambda_0 / \Lambda ) \)
\eeq
where $g_0 = g(\Lambda_0)$.    Thus at low energies where $\Lambda < \Lambda_0$,  
the combination $2\pi/mg$ grows.   However since the dependence on the cut-off is only
logarithmic,  it is not 
clear that this could explain the large $C$ discrepancy.  
This seems to imply that higher N-body interactions are important.

\section{Partition function in terms of the S-matrix}

\subsection{Formal expression for $Z$ in terms of the S-matrix}

We are interested in the partition function:
\beq
\label{con.1}
Z(\beta, \mu) = \Tr ~ e^{-\beta (H - \mu N)}
\eeq
where $\beta = 1/k_B T$ and $\mu$ is the chemical potential.
We henceforth set $k_B=1$.    The trace will be performed over 
the free particle Fock space.  Assuming only one kind of particle,
the trace is computed based on  the following resolution of the identity:
\beq
\label{con.2}
{\bf 1}  = \sum_{N=0}^\infty \inv{N!} \int  
\dk{1} \, \dk{2}  \cdots \dk{N} ~~ 
| \kvec_1 \cdots \kvec_N \rangle \langle \kvec_1 \cdots \kvec_N | 
\eeq
where $\dkonly \equiv d^d\kvec / (2\pi)^d$ and $d$ is the spatial
dimension.     
We adopt the following conventions:
\beq
\label{con.3}
|\kvec_1, \ldots, \kvec_N\rangle =  \adag_{\kvec_1} \cdots
\adag_{\kvec_N} \, |0\rangle 
\eeq

\beq
\label{con.4}
a_\kvec a^\dagger_\kvecp ~ -s ~ a^\dagger_\kvecp a_\kvec = \2pid \, 
 \delta^{(d)} (\kvec - \kvecp )
\eeq
where $s=1$ corresponds to bosons and $s=-1$ fermions. 
For simplicity we will use the following notation:
$\delta_{\kvec, \kvec'} \equiv \delta^{(d)} (\kvec - \kvec' )$.
The energy of a free one-particle state will be denoted as 
$\omega_\kvec$.  
We will also need:
\beq
\label{con.7}
\delta_{\kvec , \kvec} = \delta^{(d)} (0) = \frac{ V }{\2pid} 
\eeq
where $V$ is the volume.

As usual, we assume the hamiltonian can be split 
into a free part $H_0$
and an interacting part: $H=H_0 + H_I$. 
Let $\hat{S} (E)$ denote the S-matrix operator in the formal
theory of scattering where $E$ is an off-shell energy variable.
   It can be expressed in the conventional
manner:
\beq
\label{S.1}
\hat{S} (E) = 1 + 2 \pi i \,  \delta (E-H_0) \, \hat{T}(E)
\eeq
where $H_0$ is the free hamiltonian operator  and everywhere the over-hat
denotes a quantum operator.      
On shell the matrix elements of $\hat{T} (E)$ are 
$T=(2\pi)^d \delta_{\{\kvec , \kvec' \}} \CM_{\{\kvec\} \to \{\kvec' \} }$
where $\CM$ are the scattering amplitudes.   

\def\dLR{{{\stackrel{\leftrightarrow}{\d} }}}

Our starting point is the basic formula derived in \cite{Ma}:
\beq
\label{S.3o}
Z = Z_0 + \inv{4\pi i} \int_0^\infty dE e^{-\beta E} \Tr 
\( \hat{S}^{-1} \dLR_E \hat{S} \)
\eeq
where $X \dLR_E Y = X(\d_E Y) - (\d_E X) Y$,
and $Z_0$ is the free partition function.  
We will work with the equivalent expression\footnote{
In previous work\cite{Leclair}, 
 the correction to $Z$ was written as the sum of
two terms $Z_A$ and $Z_B$ which obscured the logarithmic structure, 
and the focus was on the $Z_A$ terms; it was incorrectly argued
that the $Z_B$ terms can be discarded.}:
\beq
\label{S.3}
Z = Z_0 + \inv{2\pi} \int dE \, \rme^{-\beta E}  \,  \Tr \, 
{\rm Im}  \,   \d_E \log \hat{S}(E)   
\eeq

Since $\log \hat{S}(E) \propto \delta(E-H_0)$,  let us define the
operator  $W$ as follows:
\beq
\label{S.4}
\Im \log \hat{S} (E) \equiv  2 \pi \delta(E-H_0) \, W(E) 
\eeq
Integrating by parts,  one then has
\beq
\label{S.5}
Z = \int dE \, \rme^{-\beta E} \, \Tr \( \delta(E-H_0) (  W(E) \beta + 1)\)
\eeq
The advantage of the above expression is that in taking the
trace,  $H_0$ can be replaced with the free particle energies
and the integral over $E$ performed.  
For simplicity of notation, define
\beq
\label{S.6}
\Uhat \equiv   W  \beta + 1
\eeq

\subsection{Cluster decomposition}

 As we will see, 
 clustering properties of the S-matrix  ensure that 
the free energy depends only on connected matrix elements 
and is proportional to the volume.  For any operator $X$, 
  the cluster decomposition
can be expressed as follows:
\beq
\label{C.1}
\langle \Psi' | X  | \Psi \rangle =
\sum_{\rm partitions}  s^p  ~ 
\langle \Psi'_1 | X  | \Psi_1\rangle_c  
\langle \Psi'_2 | X | \Psi_2\rangle_c  
\langle \Psi'_3 | X | \Psi_3\rangle_c \cdots 
\eeq
where the sum is over partitions of the state $|\Psi \rangle$ 
into clusters $|\Psi_1\rangle, |\Psi_2 \rangle, ....$.
(The number of particles in $|\Psi_i\rangle$ and 
$\langle \Psi'_i |$ is not necessarily the same.)    
The above formula essentially {\it defines} what is
meant by the connected matrix elements $\langle X  \rangle_c$.

 In order for the free energy to be extensive, 
$\Uhat$ must cluster properly, as we now describe.  
Introducing the short-hand notation $
|\bfN{12} \cdots \rangle = | \kvec_1 \kvec_2 \cdots \rangle$, 
for the operator $\Uhat$ 
the above equation reads
\barray
\label{C.2}
 \brabf{1'} \Uhat   \ketbf{1}  &=& \brabf{1'} \Uhat  \ketbf{1}_c 
\\
\nonumber
\brabf{1'2'} \Uhat  \ketbf{12} &=& \brabf{1'2'} \Uhat  \ketbf{12}_c 
+ \brabf{1'} \Uhat  \ketbf{1}_c \brabf{2'} \Uhat  \ketbf{2}_c 
+ s   \brabf{2'} \Uhat  \ketbf{1}_c \brabf{1'} \Uhat  \ketbf{2}_c
\\
\nonumber
\brabf{1'2'3'} \Uhat  \ketbf{123} &=& 
\brabf{1'2'3'} \Uhat  \ketbf{123}_c 
\\
\nonumber 
&~& ~~~~+ \[ \brabf{1'} \Uhat  \ketbf{1}_c \brabf{2'3'} \Uhat  \ketbf{23}_c 
+ {\rm permutations} \]_9
\\
\nonumber 
&~& ~~~~~~~~~~+\[ \brabf{1'} \Uhat  \ketbf{1}_c \brabf{2'} \Uhat 
 \ketbf{2}_c \brabf{3'}
 \Uhat  \ketbf{3}_c 
+ {\rm permutations} \]_6 
\earray
etc. The subscript $[*]_9$ denotes the number of terms
within the bracket, so that   
  for 3 particles there is a total of $1+9+6=16$ terms on the right hand side
of the above equation.  
The connected matrix 
elements are characterized by being proportional to a {\it single}  overall 
$\delta$-function.    

When $\{ \kvec'\}  = \{ \kvec \} $, the above cluster expansion leads to another specialized  type of
cluster expansion that is suited to the computation of the partition function. 
From eq.  (\ref{C.2}) we define $\vhat_N (\kvec_1 , ..., \kvec_N)$ as factors
that cannot be written as a product of separate functions of disjoint subsets 
of the $\kvec_i$:
\barray
\label{clusterv}
\brabf{1} \Uhat  \ketbf{1} &=& \brabf{1} \Uhat  \ketbf{1}_c = \vhat_1 (\kvec_1) 
\\ \nonumber
\brabf{12} \Uhat  \ketbf{12} &=& \vhat_2 (\kvec_1, \kvec_2 ) + \vhat_1 (\kvec_1 ) 
\vhat_1 (\kvec_2) 
\\
\nonumber
\brabf{123} \Uhat  \ketbf{123} &=& \vhat_3 (\kvec_1 , \kvec_2 , \kvec_3 ) + 
\Bigl[ \vhat_2 (\kvec_2 , \kvec_3 ) \vhat_1 (\kvec_1 ) + 
\vhat_2 (\kvec_1 , \kvec_3) \vhat_1 (\kvec_2) 
+ \vhat_2 (\kvec_1 , \kvec_2 ) \vhat_1 (\kvec_3) \Bigr]
\\ \nonumber &~& ~~~~~~~~~~~~~~~~~~~~~~~~~~~~~
+ \vhat_1 (\kvec_1 ) \vhat_1 (\kvec_2 ) \vhat_1 (\kvec_3) 
\earray
etc.  
The above definition implies that the $\vhat_N$ are sums of
terms involving the connected matrix elements of $\Uhat$. For instance: 
\barray
\label{clustereg} 
\vhat_2 (\kvec_1 , \kvec_2 ) &=& \brabf{12} \Uhat  \ketbf{12}_c + 
  s \brabf{2} \Uhat  \ketbf{1}_c \brabf{1} \Uhat  \ketbf{2}_c 
\\ \nonumber
\vhat_3 (\kvec_1 , \kvec_2 , \kvec_3 ) &=& 
\brabf{123} \Uhat  \ketbf{123}_c 
+ \Bigl[  s \brabf{2} \Uhat  \ketbf{1}_c  \brabf{13} \Uhat  \ketbf{23}_c  
     + {\rm perm.} \Bigr]_6 
\\ \nonumber
&~& ~~~~~~~  
+ \Bigl[  \brabf{3} \Uhat  \ketbf{1}_c \brabf{1} \Uhat  \ketbf{2}_c \brabf{2}
 \Uhat \ketbf{3}_c + 
{\rm perm.} \Bigr]_2 
\earray
A combinatoric argument then shows:
\beq
\label{logZ}
\log Z   =  \sum_{N=1}^\infty 
\frac{z^N}{N!} \int \dk{1} \cdots \dk{N} ~  
\rme^{-\beta \sum_{i=1}^N  \omega_i } \, 
\vhat_N (\kvec_1 , ..., \kvec_N) 
\eeq
where $z\equiv e^{\beta\mu}$ and $\omega_i = \omega_{\kvec_i}$.

All of the $\vhat$ are proportional to an overall $(2\pi)^d \delta^{(d)} (0) = V$, 
so let us factor out the volume $V$ and define:
\beq
\label{2.10}
\vhat_N (\kvec_1 , ..., \kvec_N ) 
\equiv  V   \,  w_N ( \kvec_1 , ..., \kvec_N ) 
\eeq   
The 
 free energy density is   then completely determined by the functions
$w_N$:
\beq
\label{2.11}
\free = - \inv{\beta V} \log Z = -\inv{\beta} \sum_{N=1}^\infty 
\frac{z^N}{N!} \int \dk{1} \cdots \dk{N} ~  \rme^{-\beta \sum_i  \omega_i } 
\, w_N (\kvec_1 , ..., \kvec_N) 
\eeq


\def\ebH{\Uhat}

For free particles,  the only non-zero connected matrix element 
is the 1-particle to 1-particle one:  
$\langle 2 | \ebH | 1\rangle =  \langle 2| 1\rangle$.
This still implies $\vhat_N$ is non-zero for any N:
\beq
\label{free1}
\vhat_N (1,2,.., N) = s^{N-1} (N-1)!  \, 
\langle {\bf N} \ketbf{1} \langle {\bf 1} \ketbf{2} \langle{\bf 2} \ketbf{3}
 \cdots \langle {\bf N-1} \ketbf{N} 
\eeq
Note that although $\vhat_N$ is a product of N $\delta$-functions,  it only has
 a single $\delta (0)$.     All but one $\kvec$ integral is saturated by
$\delta$-functions at each $N$ and one obtains the well-known free-particle result: 
\beq
\label{free2} 
\free_0  = \frac{s}{\beta}    \int \frac{d^d \kvec}{\2pid}  ~ \log 
\( 1 - s z e^{-\beta \omega_\kvec} \) 
\eeq

\subsection{Diagrammatic description}

\def\ebH{\Uhat}

In the interacting case, terms involving the one-particle factors
$\brabf{2} \ebH \ketbf{1}_c$ can be easily summed over as in the
free particle case.   We assume that $\ket{\kvec}$ is
a stable one-particle eigenstate of $H$ with energy $\omega_\kvec$:
\beq
\label{onepart.1}
\bra{\kvec'} \ebH \ket{\kvec}_c =  \langle \kvec' \ket{\kvec} 
=   \2pid   \delta_{\kvec, \kvec'} 
\eeq
Consider for example the $\vhat_3$ contribution coming from the second terms 
in eq. (\ref{clustereg}):
\beq
\label{onepart.2}
\log Z = .... + \frac{z^3}{3!} \int  \dk{1}  
\dk{2} \dk{3} \, \rme^{-\beta \sum_{i=1}^3  \omega_i }  \, 
\( s \brabf{2} \ebH \ketbf{1}_c \brabf{13} \ebH \ketbf{23}_c  + {\rm perm. } \) +...
\eeq
Because of the $\delta$-function,  one can do one $\kvec$ integral.  Combined with
the primary term coming from $\vhat_2$ one finds: 
\beq
\label{onepart.3}
\log Z = ... + \frac{z^2}{2} \int  \dk{1}  \dk{2} \,  \rme^{-\beta(\omega_1 + \omega_2 )} \,  
\( 1 + sz (e^{-\beta \omega_1 } + e^{-\beta \omega_2 }) \) 
~\brabf{12} \ebH \ketbf{12}_c +...
\eeq
It is easy to see that the terms in the higher $\vhat_N$ that involve
one $\vhat_2$ and  $N-1$ $\vhat_1$'s contribute additional terms in 
the parentheses of the above equation that sum up to 
$\prod_{i=1}^2 (1-sz e^{-\beta \omega_i })^{-1}$.  To summarize this
result in the most convenient way,  define  $\sum'_N$  as the sum over 
$\vhat_N$ terms minus any terms involving the 1-particle factors  $\vhat_1$.  
   Then 
\beq
\label{onepart.5} 
\free  = \free_0  -  \inv{\beta}  {\sum_{N\geq 2}}'   \inv{N!} 
\int 
\( \prod_{i=1}^N \dk{i} f_0 (\kvec_i ) \) 
~ w_N (\kvec_1, ..., \kvec_N ) 
\eeq
where $\free_0 $ is the free contribution in eq. (\ref{free2}) and 
 we have defined the free ``filling fraction'': 
\beq
\label{onepart.6}
f_0 (\kvec ) \equiv \frac{z}{e^{\beta \omega_\kvec}  - s z } 
\eeq

The functions $w_N$ have many  contributions built out of
the connected matrix elements of $\Uhat$.   Define the 
vertex function $\CV(\kvec_1', ..., \kvec_m' ; 
\kvec_1 ,  ..., \kvec_n )$ as follows:
\beq
\label{vertex.1}
\langle \kvec_1', ..., \kvec_m' | \Uhat |\kvec_1 , ..., \kvec_n \rangle_c  
= \beta  (2\pi)^d \delta_{\kvec, \kvec'}  \, 
\CV (\kvec_1' , ..., \kvec_m' ; \kvec_1 , ..., \kvec_n ) 
\eeq
The vertices $\CV$ have no temperature dependence and are essentially
on-shell matrix elements of $\log \hat{S}$.     
To lowest order in the scattering matrix $\hat{T}$,  this vertex is
just the scattering amplitude:
$\CV = \CM$.   In the sequel we will compute $\CV_2$ for the interacting
Bose gas.   
Given our present interest in non-relativistic systems,
 only the matrix elements with $n = m$ are 
 of concern here.  We highlight this fact by attaching a 
subscript $n$ to $\CV$:
\beq
\CV_n (\kvec_1' , ..., \kvec_n' ; \kvec_1 , ..., \kvec_n ) \equiv \CV (\kvec_1' , ..., \kvec_n' ; \kvec_1 , ..., \kvec_n ) 
\eeq

  Each term in the  corrections  $\free - \free_0$ can be represented graphically.
  We stress that this diagramatic expansion as defined here, 
while being analogous to the usual Feynman diagrams, is completely 
unrelated to the (finite temperature) Feynman diagram approach.
This is clear since each vertex in principle represents an exact 
zero-temperature quantity summed to all orders in perturbation theory. 
The correction $\free - \free_0$ is the sum of all 
connected ``vacuum diagrams'', 
made of oriented lines and vertices with $n$ incoming and $n$ outgoing lines attached, 
$n$ being any positive integer.  Note that there must be at least one vertex; 
i.e. the circle isn't allowed since it is already included as 
$\CF_0$.    A diagram is then evaluated 
according to the following rules  and represents a  contribution to $\CF$:
\begin{enumerate}
\item Each line carries some momentum $\kvec$.  To such a line is  associated
 a factor of $f_0(\kvec)$.
\item To each vertex, with the incoming set and the outgoing set of momenta being $\lbrace \kvec_1, \dots, \kvec_n \rbrace$ and $\lbrace \kvec_1', \dots, \kvec_n' \rbrace$ respectively, associate a factor of the vertex function $(2\pi)^d \beta \,  \CV_n (\kvec_1' , ..., \kvec_n' ; \kvec_1 , ..., \kvec_n )$.
\item Enforce momentum conservation at each vertex.
\item Integrate over all unconstrainted momenta with $d^d \kvec / (2\pi)^d$.  
\item  Divide by the symmetry factor of the diagram,  defined as the number
of permutations of the internal lines that do not change the topology
of the graph,  including relative positions.  This factor is identical to
the usual symmetry factor of a Feynman diagram.   
\item  For fermions, include one statistical factor $s=-1$ for each loop.
\item Divide the result by the volume of space $V = (2\pi)^d \delta(0)$ 
and by $\beta$.   Note that this is equivalent to dividing the space-time
volume $V\beta$ since at finite temperature, $\beta$ is the circumference
of compactified time.   

\end{enumerate}
The structure of this diagrammatic expansion is very similar to 
that in the work of Lee and Yang\cite{LeeYang},  however the vertices 
are different in the two approaches.

\begin{figure}
\includegraphics{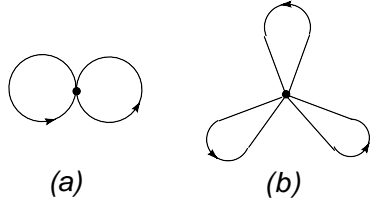}
\caption{(a) Contribution to $w_2$.  (b) Contribution to $w_3$.}
\label{w2_w3}
\end{figure}

To illustrate this diagrammatic expansion, 
we will now describe the first few terms.  
There is only one diagram that has two lines, as shown in Figure \ref{w2_w3}.  This diagram corresponds to:
\beq
\CF = .... \frac12 \int (d\kvec_1)(d\kvec_2) f_0(\kvec_1) f_0(\kvec_2) 
 \CV_2(\kvec_1, \kvec_2; \kvec_1, \kvec_2).
\eeq
Comparing  with the expression (\ref{onepart.5}), we identify
$w_2(\kvec_1, \kvec_2) = \beta \CV_2(\kvec_1, \kvec_2; \kvec_1, \kvec_2)$.
At the next order, there is again only one diagram consisting  of three lines, 
shown in Figure \ref{w2_w3},
and this implies
$w_3(\kvec_1, \kvec_2, \kvec_3) = 
 s \beta \CV_3(\kvec_1, \kvec_2, \kvec_3; \kvec_1, \kvec_2, \kvec_3)$.

\begin{figure}
\includegraphics{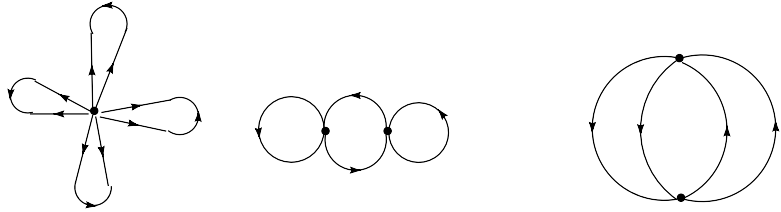}
\caption{Diagrams contributing to $w_4$}
\label{w4}
\end{figure}

There are multiple diagrams with four lines, one of them includes $\CV_4$, and the others can be built from $\CV_2$.  They are shown in Figure \ref{w4}.  Their sum is
\barray
\CF = ...+ \int &~& \left [ \prod_{i=1}^{4}(d\kvec_i) f_0(\kvec_i) \right ]
\Bigl [ \inv{4!}  \CV_4(\kvec_1, \dots \kvec_4; \kvec_1, \dots \kvec_4 )\\
\nonumber
&+& \frac{s}{2} (2\pi)^d \delta(\kvec_2 - \kvec_3) \beta \CV_2(\kvec_1, \kvec_2; \kvec_1, \kvec_3) \CV_2 (\kvec_2, \kvec_4; \kvec_3, \kvec_4) \\
\nonumber
&+& \frac{s}{8} (2\pi)^d \delta(\sum_{i=1}^{4} \kvec_i) \beta^2 \CV_2(\kvec_1, \kvec_2; \kvec_3, \kvec_4) \CV_2 (\kvec_3, \kvec_4; \kvec_1, \kvec_2) \Bigr ].
\earray
From this expression, we can identify $w_4$,  which is now a sum of terms built
from the vertices $\CV_4, \CV_2$.

An interesting class of diagrams are the ``ring-diagrams'' shown
in Figure 3,  since they can be summed up in closed form.  
Let us define
\beq
\label{rings.0}
G_2 (\kvec , \kvec ' ) = \CV_2 (\kvec, \kvec' ; \kvec, \kvec' ) 
\eeq
A ring diagram with $N$ external rings attached will have the following value according to the diagrammatic rules:
\barray
\label{single.ring.diagram}
\frac{1}{\beta N} s^{N+1} \int (d\kvec_0) \prod_{i=1}^{N}(d\kvec_i) \left
 [ \beta G_2(\kvec_0,\kvec_i) f_0(\kvec_i) \right ] f_0(\kvec_0)^N  &\qquad& \text{if $N>1$;} \\
\nonumber
\frac{1}{2 \beta } \int (d\kvec_0) (d\kvec_1) \beta 
G_2(\kvec_0,\kvec_1)  f_0(\kvec_1) f_0(\kvec_0)  &\qquad& \text{if $N=1$.}
\earray
Apart from the $N=1$ term which has a different coefficient, all other terms can be identified with the power series for the logarithmic function.  This class of diagrams 
are then resummed into:
\barray
\label{sum.ring.diagram}
\CF_{\rm ring} 
%
%
&=&
- \frac{s}{\beta} \int(d\kvec_0) \log \left (1 - s f_0(\kvec_0) \int(d\kvec_1) \beta
 G_2(\kvec_0,\kvec_1 ) f_0(\kvec_1) \right ) \\
\nonumber
&\quad& ~~~~~ \quad - \frac{1}{2} \int (d\kvec_0) (d\kvec_1)
 G_2(\kvec_0,\kvec_1) f_0(\kvec_1) f_0(\kvec_0)
\earray

\begin{figure}
\includegraphics[width=5cm]{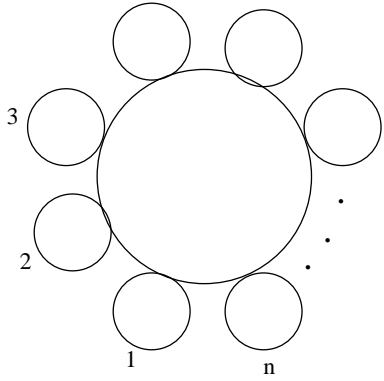}
\caption{Ring diagrams.}
\label{ring}
\end{figure}

\section{Legendre transformation and integral equation} 

\subsection{Generalities}

Given $\CF (\mu)$, one can compute
the thermally averaged number density $n$:
\beq
\label{L2}
n = - \frac{\partial \CF}{\partial \mu } \equiv  
\int \dkonly ~ f(\kvec)
\eeq 
The dimensionless
quantities $f$ are called  the filling fractions or
occupation numbers.      
One can express $\CF$ as a functional of $f$ 
 with a Legendre transformation.  Define
\beq
\label{L3}
G \equiv \CF (\mu) + \mu \,  n  
\eeq
Treating $f$ and $\mu$ as independent variables,  then 
using eq. (\ref{L2}) one has that 
$\d_\mu G =0$ which implies it can be expressed only in terms of 
$f$ and satisfies $\delta G/ \delta f = \mu$.   
Inverting the above construction shows that there exists 
a functional $\Fhat (f, \mu)$ 
\beq
\label{L4}
\Fhat (f, \mu) = G(f) - \mu \int \dkonly f(\kvec) 
\eeq
which satisfies eq. (\ref{L2}) and is a stationary point 
with respect to $f$:
\beq
\label{L5}
\frac{\delta \Fhat}{\delta f} = 0
\eeq
The above stationary condition 
 is to be viewed as determining
$f$ as a function of $\mu$.   The physical free energy is
then $\CF = \Fhat$ evaluated at the solution $f$ to the above equation.
We will refer to eq. (\ref{L5}) as the saddle point equation since
it is suggestive of a saddle point approximation to 
a functional integral.
The function $\Fhat$ is also required to satisfy
\beq
\label{Lnew.2}
\frac{  \d \Fhat }{ \d \mu }  =  - \int \dkonly \, f(\kvec)   
\eeq

In the sequel, it will be convenient to trade the chemical
potential  $\mu$ for the variable $f_0$.  We will need:
\beq
\label{Lb.1}
\frac{ \d f_0}{\d \mu }  = \beta f_0 (1 + sf_0)
\eeq
In the free
theory the functional $\Fhat_0$ is fixed  only up to
the saddle point equations,  thus its explicit expression is not unique.
As we will see, 
 the appropriate choice that 
is consistent with the diagramatic expansion is:
\beq
\label{Lnew.1}
\Fhat_0 (f, f_0) =  - \inv{\beta}  \int \dkonly 
\(  s \log ( 1 + sf)  -  \( \frac{f-f_0}{1+ s f_0} \)  \)
\eeq
(This is different than the choice made in  \cite{Leclair}.) 
The saddle point equation $\frac{ \delta \Fhat_0 }{ \delta f} =0$
implies $f=f_0$, and inserting this back into $\Fhat_0$ gives
the correct free energy $\CF_0$ for the free theory.   Furthermore,
$\Fhat_0$ satisfies eq. (\ref{Lnew.2}) if one uses the saddle
point equation $f=f_0$.

Let us now include interactions by defining
\beq
\label{var.1}
\Fhat  = \Fhat_0  + \Fhat_1   
\eeq
where $\Fhat_0$ is given in eq. (\ref{Lnew.1}) and we define
$\U$ as the ``potential'' which depends on $f$ and 
incorporates interactions:
\beq
\label{var.2}
\Fhat_1  = -\inv{\beta} 
\int \dkonly  ~ \U (f(\kvec) )
\eeq

\subsection{Integral equation}

The physical filling fraction $f$ can be obtained by differentiating 
$\free$ in (\ref{onepart.5}) with respect to the chemical potential $\mu$, 
and the result is
\beq
\label{f}
f(\kvec) = f_0(\kvec) + f_0(\kvec) \left ( s f_0(\kvec) + 1 \right ) 
\sum_{N=1}^{\infty} \inv{N!} \int  \[ \prod_{i=1}^{N} (dk_i) f_0(\kvec_i) \]  
\,  w_{N+1}(\kvec,\kvec_1,\dots,\kvec_N).
\eeq
Since $f_0$ is represented by a line in our graphical expansion, it may be tempting to postulate that $f$ would be a fully dressed line, but this interpretation isn't 
exactly correct.
Graphically, 
 a dressed line is obtained by summing over all possible insertions due to interactions.  The sum can be generated in the following way: take a (connected) vacuum diagram, cut an internal line into two external legs, and the resulting diagram contributes to the sum.  If we sum over all possible ways of cutting an internal line from a vacuum diagram, and furthermore sum over all such diagrams, we have generated all correction terms to the so-called propagator.
Cutting a line and replacing it with two legs corresponds to replacing a factor of $f_0(\kvec)$ by $s f_0(\kvec)^2$.  We have the extra factor of $s$ because cutting one line removes one loop integral.

Recall that $\free - \free_0$ is the sum of all connected vacuum diagrams.  The action of the derivative $\partial / \partial \mu$ makes sure that we've picked out every internal line in every diagram.  However, it replaces $f_0$ by $f_0(1+ s f_0)$.
Consequently, we define the dressed line $\tilde{f}(\kvec)$ by
\beq
\label{relation_f_f.tilde}
\tilde{f}(\kvec) = f_0(\kvec) + \left ( \frac{sf_0(\kvec)}{sf_0(\kvec)+1} \right ) \left ( f(\kvec) - f_0(\kvec) \right ).
\eeq
It will be useful to define $\epsilon_\kvec$ as the single particle pseudo-energy
at momentum $\kvec$ as follows:
\beq
\label{fdef}
f (\kvec) =   \inv{e^{\beta \epsilon_\kvec} -s }
\eeq
Then the eq. (\ref{relation_f_f.tilde})  can be equivalently expressed 
in the simpler form:
\beq
\label{fftilde}
\tilde{f} (\kvec)  = e^{\beta(\epsilon_\kvec  -\omega_\kvec + \mu)}  f(\kvec) 
\eeq
The following identity will be useful:
\beq
\label{df.tilde}
\frac{\delta \tilde{f}(\kvec)}{\delta f(\kvec')} = (2\pi)^d  \delta(\kvec - \kvec') \frac{sf_0(\kvec)}{1 + sf_0(\kvec)} 
\eeq

\begin{figure}
\includegraphics{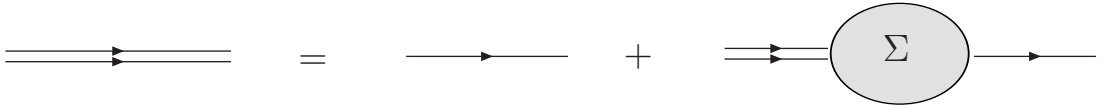}
\caption{Graphical representation of the integral equation (\ref{Dyson'}). The double line represents $\tilde{f}$, and the single line represents $f_0$.}
\label{integral equation}
\end{figure}

This fully dressed line should satisfy an  equation analogous to
 the Dyson equation for 
the full propagator:

\beq
\label{Dyson'}
\tilde{f}(\kvec) =
f_0(\kvec) +  s f_0(\kvec) \tilde{f}(\kvec) \Sigma(\kvec),
\eeq
where $\Sigma(\kvec)$ is the sum of all (amputated) one-particle (1PI) insertions. 
 Figure \ref{integral equation} represents this equation graphically.

Again we consider  the analogy to  Feynman diagrams. 
 Take a two-particle irreducible (2PI) 
 vacuum diagram with each internal line being fully dressed $\tilde{f}$, and remove one internal line to form an amputated 1PI insertion.  Summing
 over all ways of removing a line of a diagram, and then summing 
 over all 2PI vacuum diagrams, the result is $\Sigma$.

Let $- \beta \Fhat_1$ be the sum of all 2PI vacuum diagrams, with internal lines being $\tilde{f}$ instead of $f_0$.  The operations above can be neatly summarized by the following:
\beq
\label{sigma}
\Sigma(\kvec) = -s\beta \left ( \frac{sf_0(\kvec)}{1+ s f_0(\kvec)} \right )^{-1} 
\frac{\delta \, \Fhat_1 }{\delta f(\kvec)} .
\eeq
The factor of $s$ is inserted by hand because the removal of one internal line decreases the loop count by one.
One  correctly surmises  
 that $\Fhat_1$  turns out to be the interaction part of $\Fhat$.  Indeed, when there is no interaction, all vertex functions are zero and $\Fhat_1$ vanishes.
The ``Dyson''  equation now reads:
\beq
\label{Dyson}
\tilde{f}(\kvec) - f_0(\kvec) + \beta \tilde{f}(\kvec) (1+sf_0(\kvec)) 
\frac{\delta \Fhat_1 }{\delta f(\kvec)} =0.
\eeq
and  is actually an integral equation for $f$.  

If we demand that the saddle point equation of the functional $\Fhat$ leads to (\ref{Dyson}), then $\Fhat_0$ can be uniquely fixed up to an additive constant and an overall scale.  If we further demand that $\Fhat_0$ reproduces the result of the free theory, then one can show that the only consistent choice is  $\Fhat = \Fhat_0 + \Fhat_1$,  where $\Fhat_0$ is given in eq. (\ref{Lnew.1}) and
\beq
\label{eqqq}
\Fhat_1 =   \CF^{2PI} (f_0 \to \tilde{f})
\eeq
i.e. $\Fhat_1$ is given by the 2PI diagrams with $f_0$ replaced by
the fully dressed $\tilde{f}$.  
In Appendix A, we  prove that the saddle point value of $\Fhat$ 
is exactly $\free$, thus completing  the derivation.

\section{Two-body  aproximation}

\begin{figure}
\includegraphics[width=5cm]{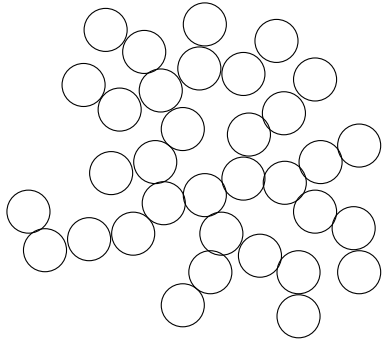}
\caption{Foam diagrams.}
\label{foam}
\end{figure}

We shall now make the following approximations:

\begin{enumerate}
\item include only the four-vertex, i.e. $\CV_2$, in our diagrammatic expansion;
\item consider only the foam diagrams, shown in Figure (\ref{foam}). 
\end{enumerate}

\noindent
This should be the
 leading contribution in the low-density, high-temperature limit.  When the density of the particles is low, 
we expect the two-particle scattering to be the most significant.  At any given order of $f_0$, a foam diagram has the least number of vertices; since each vertex contributes a factor of $\beta$, the foam diagrams dominate in the limit of small $\beta$.
In practice, this approximation amounts to setting 
\beq
\label{foam.F1}
\Fhat_1 = - \frac12 \int(d\kvec)(d\kvec') \tilde{f}(\kvec) 
\tilde{f}(\kvec') G_2 ( \kvec,\kvec'),
\eeq
and truncating all terms with more $\tilde{f}$.

Within this approximation, the integral equation (\ref{Dyson}) can be rewritten into a self-consistent equation for the single-particle pseudo-energy.  With the expressions (\ref{foam.F1}) and (\ref{df.tilde}), the equation (\ref{Dyson}) reduces to
\beq
\label{foam.Dyson}
\tilde{f}(\kvec) = f_0(\kvec) + \tilde{f}(\kvec) \left ( s \beta f_0(\kvec) \int (d\kvec') G_2 (\kvec, \kvec' )  \tilde{f}(\kvec') \right ).
\eeq
In terms of the pseudo-energy $\epsilon_\kvec$ defined in eq. 
(\ref{fdef}), using eq. (\ref{fftilde}) one can 
 rewrite (\ref{foam.Dyson}) as follows:
\beq
\label{pseudo.energy}
\epsilon_{\kvec} = \omega_{\kvec} - \mu -\frac{1}{\beta} \log \left ( 1 +
\beta  \int(d\kvec')  G_2(\kvec, \kvec')   
\frac{ e^{\beta (\epsilon_{\kvec'} - \omega_{\kvec'}+\mu )}}
{e^{\beta \epsilon_{\kvec'}} -s } \right )
\eeq

One may also want to find the free energy density of the system in terms of the physical filling fraction $f$.  By using (\ref{foam.Dyson}) and (\ref{relation_f_f.tilde}), it can be shown that
\barray
\Fhat_1 &=& - \frac12 \int(d\kvec)(d\kvec') \tilde{f}(\kvec) 
\tilde{f}(\kvec') G_2(\kvec,\kvec') \\
\nonumber
&=& - \frac{1}{2\beta} \int(d\kvec) \frac{\tilde{f}(\kvec) - f_0(\kvec) }{sf_0(\kvec)} \\
\nonumber
&=& - \frac{1}{2\beta} \int(d\kvec) \frac{f(\kvec) - f_0(\kvec) }{1 + sf_0(\kvec)}.
\earray
The free energy density  then takes the simple form:
\barray
\label{freefoam}
\free &=& \Fhat_0 + \Fhat_1 \\ 
\nonumber
&=& - \inv{\beta}  \int \dkonly 
\(  s \log ( 1 + sf)  - \inv{2}  \( \frac{f-f_0}{1+ s f_0} \)  \)
\earray

In summary,  the two-body  approximation consists of the  
integral equation (\ref{pseudo.energy}) for the pseudo-energy
defined in eq. (\ref{fdef}),  and the above expression (\ref{freefoam})
for the  free energy.  The kernel that enters this approximation
will be computed in the next section for Bose gases.

\section{Two-body kernel for interacting Bose gases}

\subsection{The kernel in arbitary dimensions.}

The interacting Bose gas in $d$ spatial dimensions 
is defined by the hamiltonian
\beq
\label{ham}
H = \int d^d \xvec \( \inv{2m} \vert \vec{\nabla} \phi (x) \vert^2 
+ 
\frac{g}{4} \vert \phi(x) \vert^4 \) 
\eeq

\def\pvec{{\bf p}}
In this section, we derive  the two-body kernel $G_2(\kvec_1, \kvec_2)$ 
that enters the approximation of the last section.  
According to the definition in Section III, 
the kernel is the connected part of the diagonal matrix element 
of $ \Im  \log \hat{S}(E)$.   Since unitarity of $S$ implies it
is a pure phase, one has 
\begin{equation}
2 \pi \delta \left (E - \frac{1}{2m}(\kvec_1^2 + \kvec_2^2) \right ) V \; G_2(\kvec_1,\kvec_2) \equiv  -i \langle \kvec_1, \kvec_2 \vert  \log\hat{S}(E) \vert \kvec_1, \kvec_2 \rangle.
\label{G2_def}
\end{equation}
where $V$ is the spatial volume. 
To define $\log (\hat{S})$, we first follow the usual convention and write the two-body $S$-matrix as the sum of the identity operator and the $T$-matrix,
as in eq. (\ref{S.1}).  The matrix elements of $\hat{T}$ are the following:
\beq
\label{Tmatrix}
\langle \pvec_1, \pvec_2 \vert \hat{T}(E) \vert \mathbf{q}_1, \mathbf{q}_2\rangle = (2 \pi)^d \delta(\mathbf{p}_1+\mathbf{p}_2 - \mathbf{q}_1 - \mathbf{q}_2) \tilde{\mathcal{M}}_{q_1,q_2 \rightarrow p_1,p_2}(E).
\eeq
where 
 $\tilde{\mathcal{M}}_{q_1,q_2 \rightarrow p_1,p_2}(E)$ denotes the two-body scattering amplitude from the in-state $\vert \mathbf{q}_1, \mathbf{q}_2\rangle$ to the out-state $\vert \mathbf{p}_1, \mathbf{p}_2 \rangle$, assuming the energy $E$ of the in-state takes on some off-shell value.  We'll denote the on-shell version by $\mathcal{M}_{q_1,q_2 \rightarrow p_1,p_2}$.
Now $\log(\hat{S})$ may be defined as a series of $\hat{T}$:
\begin{equation}
\log(\hat{S}(E)) \equiv - \sum_{n=1}^{\infty} \frac{(-1)^n}{n} \left [ 2 \pi \delta (E - H_0) \; i \, \hat{T}(E) \right ]^n .
\label{logS_series}
\end{equation}
The connectedness is guaranteed because, in terms of Feynman diagrams, $\hat{T}$ does not contain any disconnected pieces.

The diagonal matrix elements of the $n$-th term of the above operator series,
\begin{equation*}
\langle \kvec_1, \kvec_2 \vert \left [ 2 \pi \delta (E - H_0)\,  i\,  \hat{T}(E) \right ]^n \vert \kvec_1, \kvec_2 \rangle,
\end{equation*}
can be found by inserting copies of the resolution of the identity.

Note that the leftmost energy $\delta$-function  hits the state $\langle \kvec_1, \kvec_2\vert$, and consequently puts the energy $E$ on-shell.
Moreover, given that the $S$-matrix is invariant under rotations and 
Galilean boosts, the on-shell amplitude $\mathcal{M}$ can only depend on the difference of the incoming momenta, which remains invariant due to 
conservation of energy and momentum.  The amplitude $\mathcal{M}$ therefore remains constant throughout the entire available phase space, and can be taken out of any integration over the phase space.
All put together, the 
diagonal matrix element of the $n$-th term of the series \eqref{logS_series} is
\begin{equation}
\begin{split}
\langle \kvec_1, \kvec_2 \vert \left [ 2 \pi \delta (E - H_0) \, i \, \hat{T}(E) \right ]^n \vert \kvec_1, \kvec_2 \rangle
= & (2 \pi) \delta \left (E - \frac{1}{2m}(\kvec_1^2 + \kvec_2^2) \right ) V \\
& \times \left (i \mathcal{M}(\vert \kvec_1 - \kvec_2\vert)\right )^n 
\mathcal{I}^{n-1}.
\end{split}
\end{equation}
Here 
 $\mathcal{I} = \mathcal{I} (| \kvec_1 - \kvec_2|)$ 
is the volume of phase space available to the scattering out-state
resulting from summing over all two-particle states in every intermediate resolution of identity:
\begin{equation}
\mathcal{I} (|\kvec_1 - \kvec_2 |)  \equiv \frac{1}{2}\int (d\pvec_1) 
(d\pvec_2)  (2 \pi)^{d+1} \delta \left (E - \frac{1}{2m}(\pvec_1^2 + \pvec_2^2) \right ) \delta^{(d)}(\mathbf{P} - \pvec_1 - \pvec_2),
\label{phase.space.volume}
\end{equation}
where $E= (\kvec_1^2 + \kvec_2^2 )/2m$ and ${\bf P} = \kvec_1 +\kvec_2 $ 
are the total energy and momentum of the in-state.
Using this in  \eqref{logS_series}, the resummation goes through trivially,
 and one obtains:
\begin{equation}
\langle \kvec_1, \kvec_2\vert \log(\hat{S}(E)) \vert \kvec_1, \kvec_2 \rangle
= 2 \pi \delta \left (E - \frac{1}{2m}(\kvec_1^2 + \kvec_2^2) \right ) \frac{V}{\mathcal{I}} \log \left ( 1 + i \mathcal{I} \mathcal{M}(\vert \kvec_1 - \kvec_2 \vert) \right )
\end{equation}
Finally, comparing  this with \eqref{G2_def}, we have  arrived at
\begin{equation}
G_2 (\kvec_1, \kvec_2) = \frac{-i}{\mathcal{I}} \log 
\left ( 1 + i \mathcal{I} \, \mathcal{M}(\vert \kvec_1 - \kvec_2 \vert) \right ).
\label{kernel}
\end{equation}

\begin{figure}
\includegraphics[scale=0.7]{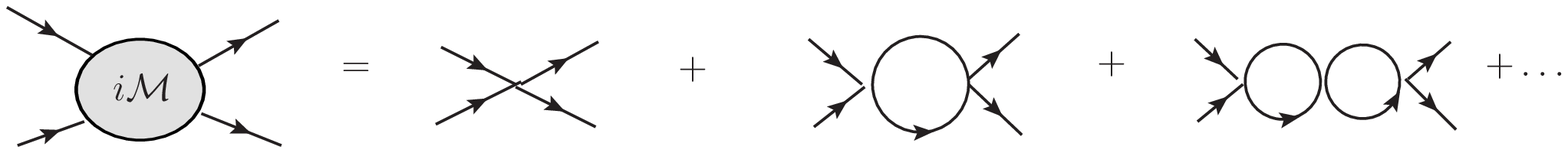}
\caption{Ladder diagram resummation of the two-body scattering amplitude.}
\label{ladder}
\end{figure}

The exact amplitude $\mathcal{M}$ in any number of dimensions $d$ 
can be calculated by resumming ladder diagrams (Fig \ref{ladder}):
\begin{align}
\begin{split}
i\mathcal{M} &= (- i g) + \dfrac{1}{2}(-ig)^2 \int(\text{loop}) + \dfrac{1}{4}(-ig)^3 \left(\int(\text{loop})\right)^2 + \dots \\
&= -ig \left ( 1 + \frac{ig}{2} \int(\text{loop}) \right )^{-1}
\end{split}
\label{exact.amplitude}
\end{align}
where the loop integral is
\beq
\label{loopint}
\int (\text{loop})  = \int (d\omega)\dkonly \left ( \frac{i}{\omega - k^2/2m + i \epsilon} \right )\;  \left ( \frac{i}{(E-\omega)- (\mathbf{P}-\kvec)^2/2m + i \epsilon} \right)
\eeq

\subsection{The kernel in two dimensions.}

The 
first step of the calculation is to compute the kernel by setting $d=2$ in equations \eqref{kernel}, \eqref{phase.space.volume} and \eqref{exact.amplitude}.
In dimension $d=2$ the phase space volume $\mathcal{I} = m/4$.

The loop integral diverges in $d=2$, and must be regularized with a UV momentum cutoff $\Lambda$.  The final form of $\mathcal{M}$ is:
\begin{equation}
\mathcal{M}(\vert \kvec_1 - \kvec_2 \vert) = - \frac{g}{1+\frac{ m g}{8 \pi}\left ( \log\left ( \frac{4 \Lambda^2}{\vert \kvec_1 - \kvec_2 \vert^2} \right ) + i \pi \right ) }
\end{equation}
When all put together, the kernel $G_2$ is the following:
\begin{equation}
\begin{split}
G_2(\kvec_1, \kvec_2) &=
\frac{-4 i}{m} \log \left ( \frac{1 + 
	\frac{mg}{8 \pi} \left (
		\log \left ( \frac{4 \Lambda^2}{\vert \kvec_1 - \kvec_2 \vert^2} \right ) - i \pi
	\right ) }
	{1  +
	\frac{mg}{8 \pi} \left (
		\log \left ( \frac{4 \Lambda^2}{\vert \kvec_1 - \kvec_2 \vert^2} \right ) + i \pi
		\right )}  \right ) \\
& =  - \dfrac{8}{m} \arctan \left( 
	\frac{mg/8}
		{1 
		+ \frac{mg}{8 \pi} 
			\log \left( \frac{4 \Lambda^2}{\vert \kvec_1 - \kvec_2 \vert^2} \right)
		}
	\right) 
\end{split}
\label{kernel.2d}
\end{equation}

\subsection{The kernel in three dimensions}

Computation of the two-body kernel in $d=3$ has little different from 
the 2-dimensional case. 
The phase space volume \eqref{phase.space.volume} turns out to be:
\beq
\mathcal{I} = \frac{m}{8 \pi} \vert \kvec_1 - \kvec_2 \vert.
\eeq
Once again the loop integral is UV divergent and in principle needs a cutoff $\Lambda$.  The divergence 
is linear in $\Lambda$, and can be completely absorbed by a redefinition of the coupling constant.  We proceed to define the renormalized coupling $g_R$ by
\beq
\frac{1}{g_R} = \frac{1}{g} + \frac{m \Lambda}{4 \pi^2}.
\eeq
The amplitude \eqref{exact.amplitude} becomes
\beq
i \mathcal{M} = \frac{- i g_R}{1 + i\frac{m g_R }{16 \pi} \vert \kvec_1 - \kvec_2 \vert }.
\eeq

The two-body kernel \eqref{kernel} is then 
\beq
\begin{split}
G_2(\kvec_1,\kvec_2) &= - i \frac{8 \pi}{m \vert \kvec_1 - \kvec_2 \vert} \log \left(
\frac{1 - \frac{i m g_R}{16 \pi} \vert \kvec_1 - \kvec_2 \vert}{1 + \frac{i m g_R}{16 \pi} \vert \kvec_1 - \kvec_2 \vert}\right) \\
&= - \frac{16 \pi}{m \vert \kvec_1 - \kvec_2 \vert} \arctan \left
 (\frac{m g_R}{16 \pi} \vert \kvec_1 - \kvec_2 \vert \right)
\end{split}
\eeq

 \section{One-dimensional case: comparison with the thermodynamic Bethe ansatz.}

In one spatial dimension, the thermodynamic Bethe ansatz (TBA) 
is applicable if the model is integrable, leading to an exact 
expression for the free energy. 
As explained in the Introduction,  our formalism is modeled after the TBA,
and is constructed from the same ingredients:  the filling fractions
and free energy are expressed in terms of a pseudo-energy $\epsilon (k)$,
and the latter satisfies an integral equation based on the exact
S-matrix.    If the model is integrable,  the n-body S-matrix factorizes
into 2-body S-matrices,  and the TBA incorporates these n-body
interactions in its derivation and the final result is expressed
only in terms of the 2-body S-matrix.    The TBA thus goes beyond the 2-body
approximation described in section V. 
Nevertheless,  our formalism should coincide with the TBA to
lowest order in the scattering kernel, and in this section we show 
this is indeed the case.

\subsection{Kernel} 

 In this section, we shall specialize in the case $d=1$.  Also, we shall set $m = 1/2$ throughout this section.
The on-shell value of energy for the two-particle in-state $\vert k_1, k_2 \rangle$ is $E = k_1^2 + k_2^2$.  With this identification, the phase space volume \eqref{phase.space.volume} turns out to be
\beq
\mathcal{I} = \frac{1}{2 | k_1 - k_2 |}
\eeq
The two-body amplitude \eqref{exact.amplitude} in $d=1$ is
\beq
i \mathcal{M} = \frac{-ig}{1+\frac{ig}{4 (k_1 - k_2)}}.
\eeq
The kernel \eqref{kernel} then becomes:
\beq
G_2(k_1, k_2) = -2i(k_1 - k_2) \log \left ( 
\frac{(k_1 - k_2) - ig/4}{(k_1 - k_2) + ig/4} \right ).
\label{kernel.1d}
\eeq

\subsection{Comparision with TBA}  

In one spatial dimenion,  the 
 interacting boson with hamiltonian 
(\ref{ham}) 
 is known to be exactly solvable using the TBA.  In particular, the free energy is
\beq
\label{TBA.free}
\free = - \frac{1}{\beta} \int (dk) \log \left ( 1 + e^{-\beta \epsilon(k)}\right ),
\eeq
where $\epsilon (k) $ obeys the Yang-Yang equation:
\beq
\label{Yang-Yang}
\epsilon(k) = k^2 - \frac{1}{\beta} \int (dk') K(k,k') \log (1+ e^{-\beta \epsilon(k')}),
\eeq
 The TBA kernel $K$ is the derivative of the $\log$ of the S-matrix
\beq
\label{smat}
S(k, k' ) =  \frac{ k - k' - i g/4}{k - k' + i g/4}
\eeq
\beq
K(k,k') = - i \d_k \log S (k,k') =  \frac{g/2}{(k-k')^2 + (g/4)^2}.
\eeq
Comparing with (\ref{free2}), one sees that the signs in (\ref{TBA.free}) are fermionic.  The TBA is really a description of the model 
 in terms  of its fermionic dual.

Expanding the right hand side of (\ref{TBA.free}) to the leading order in the kernel $K$ using (\ref{Yang-Yang}), we have
\beq
\label{TBA.expand}
\free \approx \free_0 + \int (dk) f_0(k) \left [ - \frac{1}{\beta} \int (dk') K(k , k') \log \left (1 + e^{-\beta \omega_{k'}} \right ) \right ]
\eeq
Integrating  the right hand side by parts with respect to $k'$ one obtains
\beq
\free \approx \free_0 - \inv{2}  \int (dk)(dk') f_0(k) f_0(k')
 \left ( -2i  \vert k - k' \vert \log \left( \frac{(k - k') - ig/4}{(k - k') + ig/4} \right) \right)
\eeq
The above expression 
 coincides with the $N=1$ term of \eqref{single.ring.diagram},
 the first order correction of our formulation, 
with  the kernel $G_2 (k,k')$  given by \eqref{kernel.1d}.

\section{Acknowledgments}

We would like to thank Erich Mueller for discussions.  
This work is supported by the National Science Foundation 
under grant number NSF-PHY-0757868.

\section{Appendix A}

We shall now show that the saddle point value of $\Fhat$ as defined in 
(\ref{var.1}) equals the free energy $\free$ given in (\ref{onepart.5}).  Throughout this section the integral equation (\ref{Dyson}) is always assumed to be satisfied.

First, let us define
\beq
\label{delta.f}
\Delta f(\kvec) \equiv \tilde{f}(\kvec) - f_0(\kvec),
\eeq
i.e. $\Delta f$ is the sum of all digrams with two external legs and any non-trivial insertion.  In particular, the sum is not restricted to 2PI terms only; all one-particle reducible terms are also included.  By re-arranging (\ref{relation_f_f.tilde}), we get one useful identity:
\beq
\frac{\Delta f}{f_0} = \frac{s(f-f_0)}{1+sf_0}
\label{id.delta.f}
\eeq
We will now rewrite $\Fhat_0$ in a more convenient form using (\ref{id.delta.f}):
\barray
\label{Fhat0.saddle.point}
\Fhat_0 &\equiv&
- \inv{\beta}  \int \dkonly 
\(  s \log ( 1 + sf)  -  \( \frac{f-f_0}{1+ s f_0} \)  \) \\
\nonumber
&=& - \inv{\beta}  \int \dkonly  \left [ s \log (1 + s f_0) + s \log \( 1 + \frac{\Delta f}{f_0} \) - s \frac{\Delta f}{f_0} \right ] \\
\nonumber
&=& \free_0 + \inv{\beta}  \int \dkonly \, s \sum_{n=2}^{\infty} \frac{1}{n} \( \frac{-\Delta f}{ f_0}\)^n 
\earray
As for the using interaction part of  $\Fhat$, one can show using 
the integral equation (\ref{Dyson}):
\barray
\label{Fhat1.saddle.point}
\Fhat_1 &\equiv&
-\frac{1}{2\beta} \int (d\kvec) (d\kvec') \,  \tilde{f}(\kvec) \tilde{f}(\kvec') w_2(\kvec, \kvec') \\
\nonumber
&=& -\frac{s}{2\beta} \int \dkonly \frac{\Delta f(\kvec)}{f_0(\kvec)}.
\earray
Together we conclude that the saddle point value of $\Fhat$ is
\beq
\label{Fhat.saddle.point}
\Fhat = \free_0 + \frac{s}{\beta}  \int \dkonly \, \left [ -\frac12 \frac{\Delta f}{f_0} + \sum_{n=2}^{\infty} \frac{1}{n} \( - \frac{\Delta f}{f_0}\)^n  \right ] 
\eeq

\begin{figure}
\includegraphics{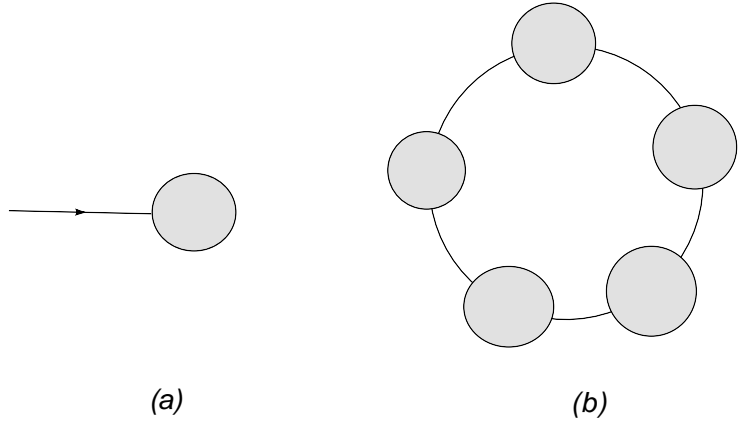}
\caption{(a) Graphical representation of $\Delta f / f_0$; one of the two external legs is amputated.  (b) The graph for $s\int \dkonly (\Delta f(k) / f_0(k))^n$.  In this example $n=5$.}
\label{line.ring}
\end{figure}

We shall now give this equation a graphical meaning.  Recall that $\Delta f$ represents the sum of all diagrams with two external legs and some non-trivial structure between the two legs.  $\Delta f / f_0$ is then the sum of such diagrams with one of the two legs amputated.  The quantity $s \int \dkonly (\Delta f(k) / f_0(k))^n$ then represents a ring with $n$ non-trivial insertions, with the factor of $s$ again coming from the loop.  (See Figure \ref{line.ring})  Again we stress that such insertion needs not be 1PI.

\begin{figure}
\includegraphics{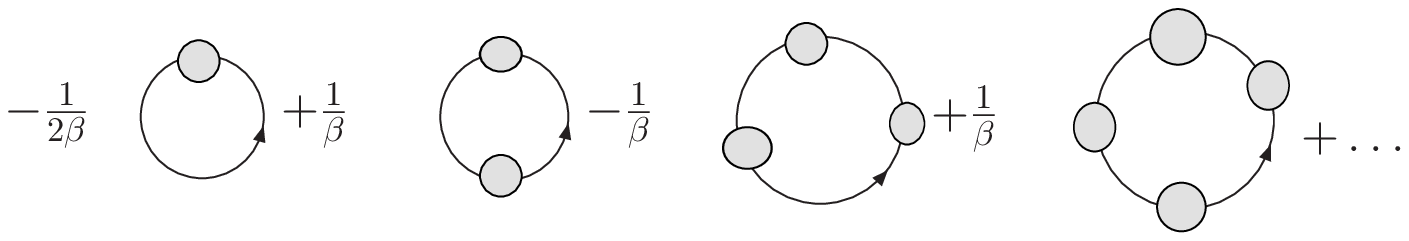}
\caption{Graphical representation of $\Fhat - \free_0$.}
\label{Fhat-free}
\end{figure}

The correction to the free energy $\Fhat - \free_0$ can then be graphically represented as Figure \ref{Fhat-free}.  The factor $1/n$ of the $n$th order term in (\ref{Fhat.saddle.point}) is cancelled because 
there are $n$ ways to order these $n$ insertions to form an identical ring, not counting possible extra symmetry.

Recall that, within the foam-diagram approximation, the free energy (\ref{onepart.5}) is really $\free_0$ plus the sum of all connected foam diagrams.  Every term in Figure \ref {Fhat-free} is a foam diagram by construction. 
  To prove the equality of (\ref{onepart.5}) and (\ref{Fhat.saddle.point}), we only need to verify (\ref{Fhat.saddle.point}), or equivalently Figure \ref{Fhat-free}, produce the correct coefficient for every foam diagram.

Consider a foam diagram with $N$ loops labeled $i = 1 \dots N$, and the $i$th loop has $n_i$ vertices attached to it.  It can be shown that
\beq
\label{sum.ni}
\sum_{i=1}^{N} n_i = 2N -2.
\eeq
Now we consider how the particular foam diagram may be constructed from Figure \ref{Fhat-free} if the loops shown is the $i$th loop of the diagram.
The $i$-th loop has $n_i$ vertices; it can be built from terms with at most $n_i$ insertions.  The $m$-th order term  contributes only if $m \leq n_i$.  The number of ways to partition the $n_i$ 1PI insertions into $m$ connected groups is simply the binomial coefficient $C^{m}_{n_i}$.  

We can now sum over every loop of the diagram to get the overall coeficient:
\barray
\label{diagram.coefficient}
\sum_{i=1}^{N} \left [ \frac12 C^{1}_{n_i} - \sum_{m=2}^{n_i}(-1)^m C^{m}_{n_i} \right ]
&=& \sum_{i=1}^{N} \left [ - \sum_{m=0}^{n_i} (-1)^m C^{m}_{n_i} - \frac12 C^1_{n_i} + C^0_{n_i} \right ] \\
\nonumber
&=& \sum_{i=1}^{N} \left [ -(1-1)^{n_i} - \frac12 n_i + 1 \right ] \\
\nonumber
&=& 1
\earray
We have used the identity (\ref{sum.ni}) to evaluate the sum.  This shows that the coefficient of a diagram without symmetry equals exactly unity.  If a diagram has an $N$-fold rotational symmetry, then the above counting argument overcounts the contribution by $N$ times, and a symmetry factor $1/N$ is needed.

We have thus proven that $\Fhat_1$ in
eq. (\ref{Fhat1.saddle.point})   generates every foam diagram with the correct coefficient.  Its saddle point value is therefore the correct free energy of the system within the foam-diagram approximation.

\end{document}